\documentstyle[12pt,axodraw,epsf,epsfig]{article}

\newcommand{\tr}{ {\rm Tr} }

\newcommand{\beq}{ \begin{eqnarray} }
\newcommand{\eeq}{ \end{eqnarray} }
\newcommand{\beqstar}{ \begin{eqnarray*} }
\newcommand{\eeqstar}{ \end{eqnarray*} }

\newcommand{\lsim}{ \mathop{}_{\textstyle \sim}^{\textstyle <} }

\newcommand{\lsp}{ \left ( }
\newcommand{\rsp}{ \right ) }

\newcommand{\llp}{ \left [ }
\newcommand{\rlp}{ \right ] }
\newcommand{\labs}{ \left | }
\newcommand{\rabs}{ \right | }

\newcommand{\GEV}{ {\rm GeV} }

\newcommand{\sla}[1]{\not\!#1}

\begin{document}
\baselineskip 0.7cm

\begin{titlepage}

\begin{center}

\hfill ICRR-Report-504-2004-2\\
\hfill TU-720\\
\hfill \today

{\large 
Hadronic EDMs induced by the strangeness \\
and  \\
constraints on supersymmetric CP phases
}
\vspace{1cm}

{\bf Junji Hisano}$^{1}$  and 
{\bf Yasuhiro Shimizu}$^{2}$
\vskip 0.15in
{\it
$^1${ICRR, University of Tokyo, Kashiwa 277-8582, Japan }\\
$^2${Department of Physics, Tohoku University, Sendai 980-8578, Japan}\\
}
\vskip 0.5in

\abstract{

We consider hadronic CP violation induced by chromoelectric dipole
moments (CEDMs) of light quarks and the QCD theta parameter
($\overline{\theta}$).  We concentrate on the strange quark CEDM
especially in this paper.  Using effective CP violating nucleon
interactions induced by the CEDMs and $\overline{\theta}$ with the
SU(3) chiral Lagrangian technique, we calculate the EDMs for the
$^{199}$Hg atom, neutron and deuteron.  We also discuss supersymmetric
contributions to the electric dipole moments with the mass insertion
approximations and give stringent constraints on the CP phases of the
flavor changing SUSY breaking terms.  }

\end{center}
\end{titlepage}
\setcounter{footnote}{0}
\section{Introduction}
The origin of CP violation in nature is a very important issue in
particle physics since CP violation is indispensable for the baryon
asymmetry in universe. In the standard model (SM), there are two CP
violating parameters; the phase of the CKM matrix and the QCD theta
$(\overline{\theta})$ parameter.  The former induces CP violation in
flavor changing processes, such as CP violating $K$ and $B$ meson decay modes
whereas the latter induces flavor conserving CP violation, such as
neutron electric dipole moment (EDM).  The experimental upperbound on
the neutron EDM gives a strong constraint on $\overline{\theta}$, as
$|\overline{\theta}| \lsim 10^{-(10-11)}$.  On the other hand, the
recent measurements of CP asymmetry in the $B$ decay modes at the Babar
and Belle experiments confirm that the phase of the CKM matrix is the
dominant source of the CP violation in the $K$ and $B$ decays.  It is
known that the phase of the CKM matrix is not much enough to explain the
baryon asymmetry in universe.  Therefore, it is very important to
search for new CP violating phases.

The minimal supersymmetric SM (MSSM) is one of the most attractive
models which describe physics beyond the SM. In the MSSM many new CP
phases may be introduced in both the flavor-diagonal and changing soft
SUSY breaking terms and these new phases can contribute to the CP violation
at low energy experiments.  Recently, the Belle collaboration
announced that CP asymmetry in $B\to
\phi K_s$ is $-0.96\pm 0.50$, which is 3.5 $\sigma$ deviation from the
SM prediction \cite{Abe:2003yt}. On the other hand, the Babar result
on the CP asymmetry in $B\to \phi K_s$ is consistent with the SM
prediction \cite{babar}.  Many papers appear to explain the anomaly in
SUSY models \cite{iroiro,Harnik:2002vs}. One of the most interesting
possibilities is that the deviation is a signal of CP violation in
right-handed bottom and strange squarks mixing. In particular, in the
SUSY SU(5) GUT with right-handed neutrinos, the large right-handed
bottom and strange squarks mixing may be induced due to the large
neutrino mixing \cite{Moroi:2000tk}.  However, we have pointed out
that the CP violation in right-handed squark mixing is strongly
constrained by the $^{199}$Hg atomic EDM through the 
chromoelectric dipole moment (CEDM) of the strange quark \cite{HS}.

Motivated by the observation that strange quark contribution to
$^{199}$Hg atomic EDM may give an important implication to other
observables, we revisit various hadronic EDMs induced by the quark
CEDMs and the QCD theta term, especially paying an attention to the
strange quark CEDM. There are several theoretical approaches to
calculate the hadronic EDMs induced by the quark CEDMs so far. In
particular, the hadronic EDMs, including the neutron EDM, are
evaluated in detail from the QCD sum rule \cite{
Pospelov:1999zx,Pospelov:1999ha,Pospelov:1999mv,Pospelov:1999rg,Pospelov:2000bw,Chan:1997fw}.
In this paper, we take a rather ``traditional'' approach based on the
chiral Lagrangian. While the QCD sum rule is a good tool to dictate
the QCD dynamics, it is difficult to incorporate the sea
quark dynamics, especially in the neutron EDM. In fact, the sigma term
in the chiral perturbation theory  suggests that the sea quark dynamics is
important in the baryon physics \cite{Zhitnitsky:1996ng}. Thus, we
adopt the chiral Lagrangian approach in this paper, and incorporate
the strange quarks in it. We derive the CP violating effective
Lagrangian with SU(3) chiral Lagrangian in the presence of the QCD
theta term and the quark CEDMs for the evaluation.

From this Lagrangian we find that the strange quark CEDM contribution
to the $^{199}$Hg EDM is induced by the $\eta^0$-$\pi^0$ mixing
originated from the isospin violation. While the strange quark CEDM
contribution is evaluated by the eta exchange contribution in the previous
paper \cite{Falk:1999tm}, the recent calculation for the Shiff
moment \cite{Dmitriev:2003kb} suggests that the contribution is
suppressed by $O(10^{-2})$. However, we show that the derived
constraint on the strange quark CEDM is comparable to the previous
result from the eta exchange contribution.  Also, for the neutron EDM, 
the loop calculation shows that the constraint on the 
strange quark CEDM is stronger than that from $^{199}$Hg EDM.  The
improvement of the deuteron EDM is proposed recently
\cite{Semertzidis:2003iq}, and the sensitivity may reach to $d_D
\sim(1-3)\times 10^{-27}e\,cm$. We discuss the sensitivity to the CP
violating parameters in QCD, including the strange quark CEDM.

We also discuss the SUSY contributions to the hadronic EDMs through
the quark CEDMs. We consider the gluino contribution with the mass
insertion approximations and show the constraints on the CP phases of
the flavor changing SUSY breaking terms, which are relevant to 
the SUSY flavor physics, from the hadronic EDM experiments.

This paper is organized as follows: In Section~\ref{sec:qcp} the CP
violation at quark level up to the dimension five operators is
reviewed.  In Section~\ref{sec:hcp} we derive the effective nucleon
interactions induced by the CP violating interaction at quark level.
In Section~\ref{sec:edm} we estimate various hadronic EDMs with the CP
violating nucleon interactions.  In Section~\ref{sec:susy} we consider SUSY
contributions to the hadronic EDMs with the mass insertion
approximation and show the constraints on the SUSY parameters.
Also we discuss the correlation between the hadronic EDM
and the CP asymmetry in $B\to \phi K_S$.
Finally, Section~\ref{sec:summary} is devoted to conclusion and
discussion.

\section{Hadronic CP violation in quark level}
\label{sec:qcp}

The CP violation in the strong interaction of the light quarks is
dictated by the following effective operators,
\begin{eqnarray}
\label{CPV_quark}
{\cal L}_{\mathrm \sla{CP}} = 
 \overline{\theta}\, \frac{\alpha_s}{8\pi} G {\widetilde G}
+\sum_{q=u,d,s} i \frac{{\widetilde d}_q}{2} \overline{q}\, g_s(G\sigma)\gamma_5 q,
\label{eff1}
\end{eqnarray}
up to the dimension five ones. Here, $G_{\mu\nu}$ is the SU(3)
gauge field strength, ${\widetilde G}_{\mu\nu}=\frac{1}{2}
\epsilon_{\mu\nu\rho\sigma}G^{\rho\sigma}$ and 
$G\sigma=G^a_{\mu\nu}\sigma^{\mu\nu}T^a$.  The first term in
Eq.~(\ref{eff1}) is the effective QCD theta term, and the second term
is for  the quark CEDMs.  The $\overline{\theta}$ parameter can be O(1)
generically, however, it is strongly constrained by the neutron EDM
experiments. One of the most elegant solution is to introduce the
Peccei-Quinn (PQ) symmetry \cite{Peccei:1977hh}, since the axion makes
$\overline{\theta}$ vanish dynamically. On the other hand, when the
quark CEDMs are non-vanishing, the theta term is induced
again even if the PQ symmetry is introduced, as pointed out by
Ref.~\cite{Bigi}.  Thus, in the evaluation of the effect of the quark
CEDMs, we need to include the QCD theta term systematically. In this
section we review the CP violating effective interactions in the
strong interaction of the light quarks up to the dimension five
operators and the role of the PQ symmetry.

We start from the case without the PQ symmetry, first. Let us take a
basis, in which the QCD theta term is vanishing, by the chiral rotation as
\begin{eqnarray}
\label{LCP}
{\cal L}_{\mathrm \sla{CP}} = 
-\sum_{q=u,d,s}m_q\, \overline{q}{i\alpha_q \gamma_5} q 
+\sum_{q=u,d,s}
 i \frac{{\widetilde d}_q}{2}\, \overline{q}\, g_s(G\sigma)\gamma_5\, q,
\end{eqnarray}
where we have used $|\overline{\theta}|\ll 1$.  The first term is
induced by the axial anomaly and $\alpha_q$'s satisfy the following
relations,
\begin{eqnarray}
\label{theta}
 \overline{\theta} = \sum_{q=u,d,s} \alpha_q.
\end{eqnarray}
Furthermore, we also impose for convenience a condition that the CP violating
tadpoles for $\pi^0$ and $\eta^0$ should vanish, $ \langle \pi^0 \labs
{\cal L}_{\mathrm \sla{CP}}\rabs 0 \rangle = \langle \eta^0 \labs
{\cal L}_{\mathrm \sla{CP}}\rabs 0 \rangle =0$. Using the PCAC
relation, these conditions and Eq.~(\ref{theta}) determine $\alpha_q$'s as
\begin{eqnarray}
\label{alpha_u}
 \alpha_u &=& 
\frac{m_d}{m_u+m_d}\left(\overline{\theta} - m_0^2 \frac{{\widetilde d}_s}{2 m_s}\right)
+m_0^2 \frac{{\widetilde d}_u -{\widetilde d}_d}{2(m_u+m_d)},
\\
\label{alpha_d}
 \alpha_d &=& 
\frac{m_u}{m_u+m_d}\left(\overline{\theta} - m_0^2 \frac{{\widetilde d}_s}{2 m_s}\right)
-m_0^2 \frac{{\widetilde d}_u -{\widetilde d}_d}{2(m_u+m_d)},
\\
\label{alpha_s}
 \alpha_s &=& 
\frac{m_um_d}{m_s(m_u+m_d)}
\left(\overline{\theta} 
- m_0^2\left( \frac{{\widetilde d}_u}{2 m_u}
+ \frac{{\widetilde d}_d}{2 m_d}
\right)\right)
+m_0^2 \frac{{\widetilde d}_s}{2m_s},
\end{eqnarray}
where $m^2_0$ is the ratio of the quark-gluon condensate to the quark
one. From the QCD sum rule, it is estimated as
\cite{mQCDSR}
\begin{eqnarray}
 m^2_0= \frac{\langle 0|\overline{q}g_s(G\sigma)q|0\rangle}{\langle 0|\overline{q}q|0\rangle}
\simeq 0.8 \, \GEV^2.
\end{eqnarray}
Notice that the CEDM of the strange quark gives sizable contributions
to $\alpha_u$ and $\alpha_d$ when
$\tilde{d}_u/m_u\sim\tilde{d}_d/m_d\sim\tilde{d}_s/m_s$.

This situation is changed when the PQ symmetry is introduced, since
$\overline{\theta}$ is promoted to a dynamical field, axion ($a$),
\begin{eqnarray}
 {\cal L} = a \frac{\alpha_s}{8\pi} G {\widetilde G}.
\end{eqnarray} 
If the quark CEDMs are vanishing, $\overline{\theta}(\equiv\langle a\rangle)$ is aligned dynamically
to the zero at the minimum of the axion potential.  However, the axion
potential is modified in the presence of the quark CEDMs \cite{Bigi},
\begin{eqnarray}
 V_{{\mathrm eff}}(a) =  K_1 a +\frac12 K a^2,
\end{eqnarray}
where
\begin{eqnarray}
K &=&
-\lim_{k\to 0}i\int d^4x e^{ikx}
\langle 0 \labs T\llp \frac{\alpha_s}{8\pi}G {\widetilde G}(x)\;
\frac{\alpha_s}{8\pi}G {\widetilde G}(0)\;
\rlp \rabs 0 \rangle,
\\
K_1 &=&
-\lim_{k\to 0}i\int d^4x e^{ikx}
\langle 0 \labs T\llp \frac{\alpha_s}{8\pi}G {\widetilde G}(x)\;
\sum_{q=u,d,s}
i \frac{{\widetilde d}_q}{2}\; \overline{q}\; g_s(G\sigma)\gamma_5\; q(0)
\;
\rlp \rabs 0 \rangle.
\end{eqnarray}
The linear term with respective to $a$, which is proportional to the
quark CEDMs, induces non-vanishing vacuum expectation value for
$a$. The matrix elements, $K$ and $K_1$, are evaluated using the
current algebra
\cite{Shifman:if,Bigi}, and they are given as 
\begin{eqnarray}
K &=&
\frac{m_um_d}{(m_u+m_d)}\langle 0|\bar{q} q|0\rangle\;,
\\
K_1 &=&
-\frac{m_um_d}{2(m_u+m_d)}\sum_{q=u,d,s} \frac{{\widetilde d}_q}{m_q}
\langle 0 \labs 
\overline{q} ig_s(G\sigma)\gamma_5 q
\rabs 0 \rangle\;.
\end{eqnarray}
Thus, the minimum of the axion potential is shifted in the presence of
the quark CEDMs and the following $\overline{\theta}$ parameter is
effectively induced as
\begin{eqnarray}
\label{ind_theta}
 \overline{\theta}_{{\mathrm ind}}(\equiv \langle a \rangle) = -\frac{K_1}{K}
= m^2_0\sum_{q=u,d,s}\frac{{\widetilde d}_q}{2m_q}.
\end{eqnarray}
Plugging Eq.~(\ref{ind_theta}) into Eqs.~(\ref{alpha_u}-\ref{alpha_s}),
it is found that $\alpha_q$'s become simpler  under the assumption of the
PQ symmetry,
\begin{eqnarray}
\label{alpha_PQ}
 \alpha_q &=& 
m^2_0 \frac{{\widetilde d}_q}{2m_q}\;\;\;\;(q=u,\;d\;,s).
\end{eqnarray}
The PQ symmetry suppresses the contribution of the strange quark CEDM
to $\alpha_u$ and $\alpha_d$. This means that the valence quarks
themselves in nucleon do not suffer from the strange quark CEDM.

\section{CP violating nucleon interactions}
\label{sec:hcp}
In the previous section we have considered the CP violating
interactions at the quark level. We need to translate those quark
level interactions into hadronic interactions in order to calculate
the hadronic EDMs.  This is a rather difficult task, and the sizable
theoretical uncertainties are expected for evaluation of hadronic
matrix elements.  Here, we consider an approach based on the chiral
Lagrangian technique. It is possible to discuss hadronic CP violation
with the effective Lagrangian in a systematical and simple way.  In
this section we derive the effective CP violating nucleon interactions
using the SU(3) chiral Lagrangian in order to incorporate the
strange quark contribution in nuclear interactions.

First, let us review the chiral Lagrangian without CP violation.
We consider the SU(3)$_L\times$SU(3)$_R$ chiral symmetry
and the effective Lagrangian is written in terms of the meson fields
\begin{eqnarray}
 M=\left(
\begin{array}{ccc}
 \frac{\pi^0}{\sqrt{2}}+\frac{\eta^0}{\sqrt{6}}& \pi^+& K^+\\
  \pi^- &  -\frac{\pi^0}{\sqrt{2}}+\frac{\eta^0}{\sqrt{6}}& K^0\\
  K^- & \overline{K^0}& -2\frac{\eta^0}{\sqrt{6}}
\end{array}
\right),
\end{eqnarray}
and the baryon fields
\begin{eqnarray}
 B=\left(
\begin{array}{ccc}
 \frac{\Sigma^0}{\sqrt{2}}+\frac{\Lambda^0}{\sqrt{6}}& \Sigma^+& p\\
  \Sigma^- &  -\frac{\Sigma^0}{\sqrt{2}}+\frac{\Lambda^0}{\sqrt{6}}& n\\
  \Xi^- & {\Xi^0}& -2\frac{\Lambda^0}{\sqrt{6}}
\end{array}
\right).
\end{eqnarray}
The effective Lagrangian invariant under SU(3)$_L\times$SU(3)$_R$
is given by
\begin{eqnarray}
 {\cal L}_0 &=&\frac{f_\pi^2}{4} \tr 
\left(\partial_\mu U \partial^\mu U \right)
+
\tr \left(\overline{B}\left(i \sla{\partial} -m\right)B\right)
\\
&+&\frac{i}{2}\tr \left(\overline{B}\gamma_\mu\left(
\xi\partial^\mu\xi^\dagger+\xi^\dagger\partial^\mu\xi\right)B\right)
+\frac{i}{2}\tr \left(\overline{B}\gamma_\mu  B\left(
\partial^\mu\xi \xi^\dagger + \partial^\mu\xi^\dagger\xi\right)\right)
\nonumber\\
&+&\frac{i}{2}(D+F)\tr \left(\overline{B}\gamma_\mu\gamma_5\left(
\xi\partial^\mu\xi^\dagger-\xi^\dagger\partial^\mu\xi\right)B\right)
\nonumber\\
&-&\frac{i}{2}(D-F)\tr \left(\overline{B}\gamma_\mu\gamma_5  B\left(
\partial^\mu\xi \xi^\dagger - \partial^\mu\xi^\dagger\xi\right)\right)
\nonumber
\end{eqnarray}
where 
\begin{eqnarray}
 \xi = \exp\left( \frac{i M}{\sqrt{2} f_\pi}\right)~~~
\end{eqnarray}
with $ U = \xi^2$. The phenomenological parameters $D$ and $F$ are
0.81 and 0.44, respectively.  The effective Lagrangian is expanded
by the derivatives, and the lowest order terms are important at the low
energy.  The terms relevant in our paper are summarized in Appendix A.

The pseudoscalars are massless due to the SU(3)$_L\times$SU(3)$_R$
symmetry.  However, the SU(3)$_L\times$SU(3)$_R$ symmetry is violated
by small quark masses and the pseudoscalars become massive. Also, the
isospin breaking effect has an important role in the EDMs of the
$^{199}$Hg atom and deuteron as shown later, since it leads to the
$\pi^0$-$\eta^0$ mixing,
\begin{eqnarray}
 \left(
\begin{array}{c}
 \pi'\\
 \eta'
\end{array}
\right) =
 \left(
\begin{array}{cc}
 \cos\theta& \sin\theta\\
 -\sin\theta& \cos\theta
\end{array}
\right)
 \left(
\begin{array}{c}
 \pi\\
 \eta
\end{array}
\right),
\end{eqnarray}
where $\pi'$ and $\eta'$ are for the mass eigenstates.  The mixing
angle is given as $\tan\theta=-\sqrt{3}(m_u-m_d)/(4m_s)\simeq 0.01$
for $m_u=5.1$ MeV, $m_d=9.3$ MeV and $m_s=175$ MeV.

Now we introduce the CP violating terms to the chiral Lagrangian.  We
need to evaluate the matrix element $ \langle B_a \pi^c \left| i {\cal
L}_{\sla{\rm CP}}\right| B_b\rangle$ in order to obtain the CP
violating interactions between nucleons and pseudoscalars induced by
the CP violating terms in Eq.~(\ref{LCP}). It is reduced by the PCAC
relation,
\begin{eqnarray}
\label{NNCP}
 \langle B_a \pi^c \left| i {\cal L}_{\sla{\rm CP}}\right| B_b\rangle
&=& \sum_{q,q'=u,d,s} \frac{i}{f_\pi}
\left[
(\alpha_q m_q+\alpha_{q'}m_{q'})
\langle B_a \left| \overline{q}'T_cq \right|B_b\rangle
\right.
\nonumber\\
&&\left.
-(\frac{{\tilde d_q}}{2}+\frac{{\tilde d_{q'}}}{2})
\langle B_a \left| \overline{q}'g_s (G\sigma)T_cq \right|B_b\rangle
\right],
\end{eqnarray}
where $T_c$ is a generator for the flavor SU(3) symmetry.

The matrix elements of the scalar operators in the right-handed side
in Eq.~(\ref{NNCP}) are represented as
\begin{eqnarray}
\label{eq:xy}
 \langle B_a \left| \overline{q}T_cq \right|B_b\rangle
&\equiv& X \tr \left(\overline{B_a}B_b T_c\right)
+ Y \tr\left(\overline{B_a}T_c B_b\right),
\\
\label{eq:z}
 \langle B_a \left| \overline{q}~{\bf 1}~ q \right|B_b\rangle
&\equiv& \delta_{ab} Z,
\end{eqnarray}
from consideration of the transformation property under
SU(3)$_L\times$SU(3)$_R$.  Here, ${\bf 1}$ is the ($3\times3$) unit matrix.  
The phenomenological parameters $X$, $Y$ and $Z$ can be determined by the
baryon mass splittings and the nucleon sigma
term \cite{Zhitnitsky:1996ng}.
\begin{eqnarray}
 X&=& \frac{1}{2m_s}\left(-2m_{\Xi}+3m_\Lambda-m_\Sigma\right)
=-\langle \overline{d}d \rangle+\langle \overline{s}s \rangle,
\\
 Y&=& \frac{1}{m_s}\left(m_{\Xi}-m_\Sigma\right)
=-\langle \overline{d}d \rangle+\langle \overline{u}u \rangle,
\\
 Z&=& \frac{3\sigma}{m_u+m_d}-\frac{3}{2}\frac{m_{\Xi}-m_{\Lambda}}{m_s}
=\langle \overline{u}u \rangle+\langle \overline{d}d \rangle
+\langle \overline{s}s \rangle,
\end{eqnarray}
where $\langle\overline{q}q \rangle\equiv\langle p|\overline{q}q |p\rangle$
and $\sigma$ represents the nucleon sigma term, which is estimated as
$\sigma \simeq 45$ MeV. Numerically they are given by
\begin{eqnarray}
 &&X= -1.35, ~ Y=0.72, ~ Z=7.7,
\\
&&  \langle\overline{u}u \rangle =3.5,~
\langle\overline{d}d \rangle =2.8,~
\langle\overline{s}s \rangle =1.4.
\label{vev}
\end{eqnarray}

For the evaluation of $\langle B_a \left|\overline{q}'g_s (G\sigma)T_cq
\right|B_b\rangle$, we use the following relation adopted in 
Ref.~\cite{Falk:1999tm}, which is inspired by the QCD sum rule,
\begin{eqnarray}
\label{eq:pospelov}
 \langle B_a \left| \overline{q} g_s (G\sigma)q
\right|B_b\rangle \simeq \frac{5}{3} m_0^2
\langle B_a \left| \overline{q}q\right|B_b\rangle
\end{eqnarray}
for each quark. From Eqs. (\ref{eq:xy}) and (\ref{eq:pospelov}) 
\begin{eqnarray}
\label{xy3}
 \langle B_a \left| \overline{q}'g_s (G\sigma)T_cq \right|B_b\rangle
&\equiv& \frac{5}{3} m_0^2 \left(X \tr \left(\overline{B_a}B_b T_c\right)
+ Y \tr\left(\overline{B_a}T_c B_b\right)\right).
\end{eqnarray}
This relation is, off course, approximate one, however, it makes the
formula simple as  shown now.

From the above consideration, Eq.~(\ref{NNCP}) can be reduced as
\begin{eqnarray}
 \langle B_a \pi^c \left| i {\cal L}_{\sla{\rm CP}}\right| B_b\rangle
&=&\frac{i}{f_\pi}
\langle B_a \left| \overline{q}\{ T_c, A\}
q \right|B_b\rangle.
\label{matrixa}
\end{eqnarray}
The matrix $A$ in Eq.~(\ref{matrixa}) is a diagonal matrix, $ A =
diag(A_u,~A_d,~A_s)$, with the components
\begin{eqnarray}
A_q&=& \alpha_q m_q-\frac{{\tilde d}_q}{2}\,\frac{5}{3}m_0^2.
\end{eqnarray}
Therefore, the CP violating Lagrangian is obtained in terms of the
baryon and meson fields,
\begin{eqnarray}
\label{NNCP2}
 {\cal L}_{\rm {\sla CP}}
&=&
\frac{1}{\sqrt{2} f_\pi}
\left(
X \tr \left(\overline{B}B\{M,A\}\right) 
+
Y \tr \left(\overline{B}\{M,A\}B\right)
\right.
\nonumber\\
&+&
 \left.
\frac{2}{3}(Z-X-Y)\tr\left( AM \right)\tr\left(\overline{B}B \right)
\right).
\end{eqnarray}
The relevant terms to our discussion in Eq.~(\ref{NNCP2}) are
summarized in Appendix B.

The matrix elements $A_q$'s are written in terms of the quark CEDMs
and the QCD theta parameter.  When there is no PQ symmetry, they are
given by 
\begin{eqnarray}
 A_u &=& 3.3 \times 10^{-3}\overline{\theta}-0.53 {\tilde d}_u
-0.14 {\tilde d}_d
-7.5\times 10^{-3} {\tilde d}_s~~{\mathrm GeV},
\\
 A_d &=& 3.3 \times 10^{-3}\overline{\theta}-0.26 {\tilde d}_u
-0.41 {\tilde d}_d
-7.5\times 10^{-3} {\tilde d}_s~~{\mathrm GeV},
\\
 A_s &=& 3.3 \times 10^{-3}\overline{\theta}-0.26 {\tilde d}_u
-0.14 {\tilde d}_d
-0.27 {\tilde d}_s~~{\mathrm GeV}.
\end{eqnarray}
Here, ${\tilde d}_q$'s are written in GeV unit.
On the other hand, when the PQ symmetry is introduced, $A_q$'s are 
written in terms of the quark CEDMs and become rather simple,
\begin{eqnarray}
 A_q &=& - 0.27 {\tilde d}_q~~{\mathrm GeV},\;\;\;\;(q=u,\;d\;,s).
\end{eqnarray}

\section{Hadronic EDMs}
\label{sec:edm}
In this section we consider the hadronic CP violation.  Among various
observables, the EDMs of neutron or atoms are very sensitive to the
flavor conserving CP violation.  In particular, the experimental upper
bound on the EDMs of neutron \cite{Harris:jx} and $^{199}$Hg
atom \cite{Romalis:2000mg}, 
\begin{eqnarray}
\label{dhbound}
 |d_{\rm Hg}| < 2.1\times 10^{-28} e\, cm,
\\
\label{dnbound}
 |d_{n}| < 6.3 \times 10^{-26} e\, cm,
\end{eqnarray}
give strong constraints on hadronic CP violation. Also, the
improvement of the deuteron EDM is proposed recently, and the
sensitivity may reach to $d_D \sim(1-3)\times 10^{-27}e\,cm$
\cite{Semertzidis:2003iq}. Therefore, we estimate the EDMs of
$^{199}$Hg atom, neutron and deuteron with the chiral Lagrangian
obtained in the previous section.

\subsection{$^{199}$Hg Atomic EDM}
The $^{199}$Hg atom is a diamagnetic atom, in which electrons make a
close shell.  In such atoms, the atomic EDMs are primary sensitive to
the CP violation in nucleons and represented by the Shiff moments
$(S)$.  The Shiff moment generates the T odd electrostatic
potential $V_{\rm eff}= 4\pi S
\,(\vec{I}\cdot\vec{\nabla})\delta(\vec{r})$ with
the nucleus of spin $(\vec{I})$. The $^{199}$Hg atomic EDM is
calculated in terms of the Shiff moment $S$
\cite{Dzuba:2002kx}
\begin{eqnarray}
 d_{\rm Hg}=- 2.8\times 10^{-17}\,S({\rm ^{199}Hg})\; e\;cm,
\end{eqnarray}
where $S$ is in unit of  $e\;{fm}^{3}$. 
This implies that the Shiff moment is bounded as \cite{Dmitriev:2003sc}
\begin{eqnarray}
|S|&<& 0.75 \times 10^{-11}\; e\;{fm^3},
\end{eqnarray}
from Eq.~(\ref{dhbound}).

The Shiff momentum is induced by the pion and eta exchanges with the CP
violating coupling. From the formula derived in the previous section,
the CP violating pion interactions are given as 
\begin{eqnarray}
 {\cal L}_{\rm {\sla CP}}^{NN\pi}
&=&
\frac{1}{2 f_\pi}
\left(\left(A_u+A_d\right)
-
\frac{\theta}{3\sqrt{3}}\left(A_u-A_d\right)
\right)
\left(\langle\overline{u}u\rangle -\langle \overline{d}d \rangle 
\right)\overline{N} {\tau^a} N \pi^a
\nonumber\\
&&+
\frac{1}{2f_\pi}
\left(
\left(\left(A_u-A_d\right)-
\frac{\theta}{\sqrt{3}}\left(A_u+A_d\right)
\right)
\left(\langle\overline{u}u\rangle +\langle \overline{d}d \rangle 
\right)
+
\frac{4\theta}{\sqrt{3}} A_s \langle\overline{s}s\rangle
\right)
\overline{N} N \pi^0
\nonumber\\
&&
+
\frac{\theta}{6\sqrt{3}f_\pi}\left(A_u-A_d\right)
\left(\langle\overline{u}u\rangle -\langle \overline{d}d \rangle 
\right)
(\overline{N}\tau^a  N \pi^a-3 N\tau^3N\pi^0)
\nonumber\\
&\equiv& \overline{g}_{\pi NN}^{(0)}\overline{N} \tau^a  N \pi^a
   +\overline{g}_{\pi NN}^{(1)}\overline{N}  N \pi^0
   +\overline{g}_{\pi NN}^{(2)}(\overline{N}\tau^a  N \pi^a-3 N\tau^3N\pi^0),
\label{picp}
\end{eqnarray}
where $N=(p,n)^T$, and $\overline{g}_{\pi NN}^{(0)}$,
$\overline{g}_{\pi NN}^{(1)}$ and $\overline{g}_{\pi NN}^{(2)}$
correspond to the isoscalar, isovector and isotensor coupling
constants, respectively. Here, we include the $\pi^0$-$\eta^0$ mixing.
The CP violating eta ones are
\begin{eqnarray}
 {\cal L}_{\rm \sla{CP}}^{NN\eta} &=& \frac{1}{\sqrt{3}f_\pi}\left(
\frac{1}{2}\left(A_u+A_d\right)
\left(\langle \overline{u}u \rangle+\langle \overline{d}d \rangle\right)
-2  A_s \langle \overline{s}s \rangle
\right) \overline{N}N\eta
\\
&&+\frac{1}{2\sqrt{3}f_\pi}
\left(A_u-A_d\right)
\left(\langle \overline{u}u \rangle-\langle \overline{d}d \rangle\right)
 \overline{N}\tau_3 N\eta
\nonumber\\
&\equiv& \overline{g}^{(0)}_{\eta NN}\overline{N}N\eta
+\overline{g}^{(1)}_{\eta NN}\overline{N}\tau_3 N\eta.
\end{eqnarray}

The recent evaluation for the Shiff moment of $^{199}$Hg
\cite{Dmitriev:2003kb}, in which the many-body treatment is performed,
reveals that the core polarization effect reduces the isoscalar and
tensor channel contributions while the isovector channel one is
comparable to the previous results \cite{Flambaum:2001gq}. Their
result for the Shiff momentum is\footnote{
While only pion exchange is included in in their calculation, they
show that the correction from the finite range of the pion interaction
is small.  Thus, the above introduction of the eta contribution is
justified.  
}
\begin{eqnarray}
S({\rm ^{199}Hg})&=&
-4\times 10^{-4}
\times
(g_{\pi NN}^{(0)}
\overline{g}_{\pi NN}^{(0)}
-\frac13\frac{m_\pi^2}{m_\eta^2}
g_{\eta NN}^{(0)}
\overline{g}_{\eta NN}^{(0)})
\nonumber\\
&&
-5.5\times 10^{-2}\times
(g_{\pi NN}^{(0)}
\overline{g}_{\pi NN}^{(1)}
-\frac{m_\pi^2}{m_\eta^2}
g_{\eta NN}^{(0)}\overline{g}_{\eta NN}^{(1)})
\nonumber\\
&&+9
\times 10^{-3} \times
g_{\pi NN}^{(0)}
\overline{g}_{\pi NN}^{(2)}
\;\;  [e\;{fm^3}].
\label{shiffmom}
\end{eqnarray}
Here, $g_{\pi NN}^{(0)}$ and $g_{\eta NN}^{(0)}$ is the CP even pion
and eta coupling constants, respectively,\footnote{
The isovector coupling in the CP even pion interaction is also induced
by the $\eta^0$-$\pi^0$ mixing, but we neglect it, since it does not change
our qualitative discussion. }
\begin{eqnarray}
 {\cal L}_{\rm CP} &=& 
\frac{m_p}{f_\pi}(D+F)\overline{N}\gamma_5 \tau^a N \pi^{a'} 
-\frac{m_p}{\sqrt{3}f_\pi}(D-3F)\overline{N}\gamma_5 N \eta^{0'} 
\\
&\equiv&
g_{\pi NN}^{(0)}\overline{N}\gamma_5 \tau^a N
\pi^{a'}
+
g_{\eta NN}^{(0)}\overline{N}\gamma_5 N
\eta^{0'}.
\end{eqnarray}
The results in Ref.~\cite{Flambaum:2001gq}, which neglect the core
polarization, correspond to 
$ S({\rm ^{199}Hg})= -0.086\times g_{\pi NN}^{(0)}(
\overline{g}_{\pi NN}^{(0)}
+\overline{g}_{\pi NN}^{(1)}
-2 \overline{g}_{\pi NN}^{(2)}
)\;\; e\;{fm^3}$ for the pion contribution \cite{Dmitriev:2003kb}.

From this evaluation, it is found that the EDM of $^{199}$Hg atom is
sensitive to $(A_u-A_d)$ through the isovector channel, and the
contribution proportional to $(A_u+A_d)$ is suppressed by $O(10^{-2})$
due to the $\eta^0$-$\pi^0$ mixing or the isoscalar channel. This
means that the $^{199}$Hg atomic EDM is insensitive to the QCD theta
parameter.  Also, notice that $\overline{g}_{\pi NN}^{(1)}$ depends on
$A_s$ through $\pi^0$-$\eta^0$ mixing. While eta itself has the CP
violating interaction which depends on $A_s$, the eta exchange
contributions are suppressed due to the isoscalar channel or the mass.
Thus we concentrate on the isovector channel of the pion exchange from
the CP violating pion interaction with non-vanishing $\eta^0$-$\pi^0$
mixing.

The Shiff moment from non-vanishing $\overline{g}_{\pi NN}^{(1)}$ 
is represented in terms of $A_q$'s as  
\begin{eqnarray}
S({\rm ^{199}Hg})=-(23\times (A_u-A_d)+0.13\times A_s)\;\;  e\;{fm^3},
\end{eqnarray}
where $A_q(q=u,d,s)$ are given in GeV unit. The current experimental
bound on the EDM of $^{199}$Hg atom implies that
\begin{eqnarray}
e|\tilde{d}_u-\tilde{d}_d+0.0051 \tilde{d}_s|<2.4\times 10^{-26}e\;cm.
\label{hgbound}
\end{eqnarray}
This is almost independent of whether we impose the PQ symmetry or not,
since it depends on $A_u-A_d$. The strange quark CEDM is bounded as
\begin{eqnarray}
e|\tilde{d}_s|<4.7\times 10^{-24}e\;cm.
\label{ds}
\end{eqnarray}

It is found that this bound does not have a qualitative difference,
especially in the contribution of the strange quark CEDM, from one
based on Ref.~\cite{Flambaum:2001gq}, that is,
$e|\tilde{d}_u-\tilde{d}_d+0.012
\tilde{d}_s|<7\times 10^{-27}e\;cm$ \cite{Falk:1999tm}. In our  evaluation
the contribution of the strange quark CEDM comes from the $\eta^0$-$\pi^0$ mixing, and it is suppressed by $\sim (m_u-m_d)/(m_s m_\pi^2)$.  On the
other hand, the eta exchange contribution in the previous evaluation is
suppressed by $\sim 1/m_\eta^2$. Thus, it is found that they are
comparable.

\subsection{Neutron EDM}

The evaluation of the neutron EDM has a long history.
However, the precious calculation is still a very difficult issue.
One of the simple estimations is based on the Naive Dimensional Analysis
\cite{Manohar:1983md}. However, we can know only the order of magnitude
of the neutron EDM in this method. More elaborated calculations are
done based on the QCD sum rule \cite{
Pospelov:1999zx,Pospelov:1999ha,Pospelov:1999mv,Pospelov:1999rg,Pospelov:2000bw,Chan:1997fw}.
The QCD sum rule is an established technique, and it has a merit for
the evaluation of the neutron EDM. In
Ref.~\cite{Pospelov:2000bw} the QCD sum rule analysis reproduces the
ratio of the quark EDM contributions in the neutron EDM, which is
expected from the non-relativistic quark model. On the other hand,
they argue that the PQ symmetry suppresses the strange quark CEDM
contribution in the neutron EDM.  However, the suppression is for 
the valence quarks, not for the sea quarks. They introduce the
contribution of the strange quark CEDM only through the vacuum
expectation value of $\bar{q}i \gamma_5 q$ with $q$ the valence quark under
the CP violating background. It should be suppressed under the PQ
symmetry as discussed in Section~2.  In fact, the sigma term in the
chiral perturbation theory suggests that the sea quark dynamics is
important in the baryon physics \cite{Zhitnitsky:1996ng}. As shown in
Eq.~(\ref{vev}), the matrix element of the strange quark 
is comparable to those of the other light quarks in nucleon.

Thus, we evaluate the neutron EDM from the effective Lagrangian
obtained in the previous section by the traditional loop calculation
so that we include the strange quark contribution.  The neutron EDM
$(d_n)$ is defined by
\begin{eqnarray}
 {\cal L}_{\rm EDM} = -   \frac{d_n}{2} \overline{n} \sigma_{\mu\nu} \gamma_5 n
F^{\mu\nu}.
\end{eqnarray}
The EDM is  induced by the one loop diagrams with the charged mesons 
in Fig.~1 and they are written as
\begin{eqnarray}
  d_n &=& \frac{e}{4\pi^2f_\pi^2}
\left( A_u x^{(n)}_u + A_d x^{(n)}_d + A_s x^{(n)}_s\right),
\end{eqnarray}
where
\begin{eqnarray}
 x_u^{(n)} &=& (D+F)\left(\langle \overline{u}u \rangle
-\langle \overline{d}d \rangle\right)\log\frac{m_p}{m_\pi}
+(D-F)\left(\langle \overline{d}d \rangle
-\langle \overline{s}s \rangle\right)\log\frac{m_p}{m_K}=2.0,
\\
 x_d^{(n)} &=& (D+F)\left(\langle \overline{u}u \rangle
-\langle \overline{d}d \rangle\right)\log\frac{m_p}{m_\pi}=1.7,
\\
 x_s^{(n)} &=& (D-F)\left(\langle \overline{d}d \rangle
-\langle \overline{s}s \rangle\right)\log\frac{m_p}{m_K}=0.33.
\end{eqnarray}
Here, we have considered the most singular terms, which have
logarithmic singularities in the chiral limit
originated from $\pi$ and $K$ mesons.  

Let us consider the constraint from the neutron EDM experiment.
When there is no PQ symmetry, the neutron EDM is given 
in terms of the quark CEDMs and the QCD theta parameter,
\begin{eqnarray}
 d_n =( 7.7 \times 10^{-16} \overline{\theta}
-4.6\times {\tilde d}_u 
-3.0\times {\tilde d}_d 
+0.33\times {\tilde d}_s)   ~{e\; cm}.
\end{eqnarray}
If we impose the PQ symmetry, 
\begin{eqnarray}
 d_n =( 
1.6\times {\tilde d}_u 
+1.3\times {\tilde d}_d 
+0.26\times {\tilde d}_s)    ~{e\;cm}.
\end{eqnarray}
Here, the quark CEDMs in the above equations are written in $cm$ unit.
For comparison to the result of the QCD sum rule in
Ref.~\cite{Pospelov:2000bw}, the contributions from the up and down CEDMs 
are slightly larger in our evaluation. It is reasonable since
the sea quarks also contribute to the neutron EDM in addition to 
the valence quarks.

From the current experimental bound of the neutron EDM in 
Eq.~(\ref{dnbound}), we obtain the following bounds for the the
theta parameter and the quark CEDMs for a case that the PQ symmetry
does not exist,
\begin{eqnarray}
|\overline{\theta}|&<&8.2 \times 10^{-11}, \nonumber\\
e|\tilde{d}_u|&<&1.4\times 10^{-26}\;e\;cm, \nonumber\\
e|\tilde{d}_d|&<&2.1\times 10^{-26}\;e\;cm,\nonumber\\
e|\tilde{d}_s|&<&1.9\times 10^{-25}\;e\;cm.
\label{nedm_nopq}
\end{eqnarray}
If the PQ symmetry works, 
\begin{eqnarray}
e|\tilde{d}_u|&<&3.9\times 10^{-26}\;e\;cm, \nonumber\\
e|\tilde{d}_d|&<&4.8\times 10^{-26}\;e\;cm,\nonumber\\
e|\tilde{d}_s|&<&2.4\times 10^{-25}\;e\;cm.
\label{nedm_pq}
\end{eqnarray}
Here, we assume no accidental cancellation among the various
contributions. 

Notice that the constraint on the strange quark CEDM is one order of
magnitude stronger than that from the $^{199}$Hg atomic EDM, though the
prediction for the neutron EDM may be expected to have more
theoretical uncertainty.  While the experimental bound on the neutron
EDM is weaker than that of the $^{199}$Hg atomic EDM, the strange quark
constraint is not suppressed by the $\eta^0$ mass or the isospin
violation.  

\subsection{Deuteron EDM}

The new measurement for the deuteron EDM is proposed in 
Ref.~\cite{Semertzidis:2003iq}, and it is argued
that one or two orders of magnitude improvement may be achieved relative
to the current bounds on the QCD theta parameter and the nuclear
force. It may give a stringent constraint on the new physics or
discover the signal. Also, the nuclear dynamics in deuteron is rather
transparent and the theoretical uncertainty is expected to be small.
The deuteron EDM is given by
\begin{eqnarray}
d_D&=&(d_n+d_p)+d^{NN}_D,
\end{eqnarray}
where $d_p$ is the proton EDM and the second term comes from the CP
violating nuclear force.

In Ref.~\cite{Lebedev} they show that the chiral logarithms in the sum
of the proton and neutron EDM exactly cancel in the chiral
perturbation theory, and they adopt the QCD sum rule for it. However,
the cancellation in the chiral perturbation theory does not come from
some symmetry. The chiral logarithms in the neutron and proton EDMs
come from loop diagrams of charged mesons in which photons are
attached to the changed mesons.  Thus, it is obvious that the chiral
logarithms are canceled in the SU(2) chiral perturbation
theory. However, it is not true if we introduce the strange quark and
the corresponding mesons. 

In the SU(3) chiral perturbation theory, the sum of the proton and
neutron EDMs is
\begin{eqnarray}
(d_p+d_n) &=& \frac{e}{4\pi^2f_\pi^2}
\left( A_u x^{(n+p)}_u + A_d x^{(n+p)}_d + A_s x^{(n+p)}_s\right),
\end{eqnarray}
where
\begin{eqnarray}
 x_u^{(n+p)} &=& -\frac13\left(
D(
\langle \overline{u}u \rangle
-5\langle \overline{d}d \rangle
+4\langle \overline{s}s \rangle)
+3F(
\langle \overline{u}u \rangle
+\langle \overline{d}d \rangle
-2\langle \overline{s}s \rangle)
\right)\log\frac{m_p}{m_K} 
\nonumber\\
&&
+\frac 13 D(
6 \langle \overline{u}u \rangle
-5 \langle \overline{d}d \rangle
+\langle \overline{s}s \rangle
)
+F(
4\langle \overline{u}u \rangle
-\langle \overline{d}d \rangle
-\langle \overline{s}s \rangle
),
\\
 x_d^{(n+p)} &=& 
\frac13 D(
3\langle \overline{u}u \rangle
-10 \langle \overline{d}d \rangle
+3\langle \overline{s}s \rangle
)
+ F(
\langle \overline{u}u \rangle
-\langle \overline{s}s \rangle
)
\\
 x_s^{(n+p)} &=& 
-\frac13\left(
D(
\langle \overline{u}u \rangle
-5\langle \overline{d}d \rangle
+4\langle \overline{s}s \rangle)
+3F(
\langle \overline{u}u \rangle
+\langle \overline{d}d \rangle
-2\langle \overline{s}s \rangle)
\right)\log\frac{m_p}{m_K} 
\nonumber\\&&
+\frac13 D(
\langle \overline{u}u \rangle
-5 \langle \overline{d}d \rangle
+6\langle \overline{s}s \rangle
)
+F(
\langle \overline{u}u \rangle
+ \langle \overline{d}d \rangle
-4\langle \overline{s}s \rangle
)
\end{eqnarray}
and then $x_u^{(n+p)}=6.4$, $x_d^{(n+p)}=-2.7$, and 
$x_s^{(n+p)}=-0.40$. The logarithmic terms in the proton EDM come
from the $K^+$-$\Lambda^0$ loop in addition to the $K^+$-$\Sigma^-$
and $\pi^+$-$n$ loops. We also include constant terms which 
are not suppressed by proton mass. 
The strangeness has an important role in the
deuteron EDM.  When the PQ symmetry does not exist,
\begin{eqnarray}
(d_p+d_n) &=& 
(6.4\times 10^{-16}\overline{\theta}
-7.7\times \tilde{d}_u
+0.73\times \tilde{d}_d
+0.23\times \tilde{d}_s)\;e\;cm.
\end{eqnarray}
If it does,
\begin{eqnarray}
(d_p+d_n) &=& 
(-5.1\times \tilde{d}_u
+2.1\times \tilde{d}_d
+0.32\times \tilde{d}_s)\;e\;cm.
\end{eqnarray}

In Ref.~\cite{Khriplovich:1999qr} it is shown that $d^{NN}_D$ comes
from the isovector coupling of pion $\overline{g}_{\pi NN}^{(1)}$,
given in Eq.~(\ref{picp}), as
\begin{eqnarray}
d^{NN}_D&=&-2.1\times 10^{-15}\times 
{g}_{\pi NN}^{(0)}
\overline{g}_{\pi NN}^{(1)}\;e\;cm.
\end{eqnarray}
It implies 
\begin{eqnarray}
d^{\pi NN}_D&=&
(
-11\times \tilde{d}_u
+11\times \tilde{d}_d
-0.063\times \tilde{d}_s)\;e\;cm.
\label{edm_d}
\end{eqnarray}

From the above evaluation, the measurement of the deuteron EDM will be
a stringent test for the SM. If they establish the sensitivity of $d_D
\sim(1-3)\times 10^{-27}e\,cm$, they can probe $e\tilde{d}_u\sim
e\tilde{d}_d\sim 10^{-28}\;e\;cm$ from the nuclear force
as shown in Ref.~\cite{Lebedev}, and $\overline{\theta}\sim 10^{-11}~e\;cm$ from the
nucleon EDMs. Also, for the strange quark CEDM, they may reach to
$e\tilde{d}_s\sim 10^{-26}\;e\;cm$ from the nucleon EDMs.

\section{SUSY contributions}
\label{sec:susy}
\subsection{Hadronic EDMs}
\label{sec:susyedm}
Let us consider the SUSY contributions to the hadronic EDMs.  In the
MSSM many CP violating parameters can be introduced in the soft SUSY
breaking terms.  If there are SUSY CP phases, the quark CEDMs are
induced through one loop diagrams. The bounds on the SUSY
contributions to the hadronic EDMs can be estimated from the results
in Section~\ref{sec:edm}.

In our previous paper \cite{HS}, we have considered the gluino
contribution to the strange quark CEDM induced by the flavor mixing in
the MSSM. While the EDMs are flavor conserving phenomena, the stringent
bounds for the EDMs give constraints on even the flavor mixing, and
they have an impact on the flavor physics, such as $B$ meson decay. Thus, we
consider the gluino contributions including the flavor violation
inside the Feynman graph in order to demonstrate the constraints on
the SUSY CP phases from the our calculations.\footnote{
In
Ref.~\cite{EKY}, it is pointed out that the chargino diagrams can also
give large contribution to $^{199}$Hg EDM.  A comprehensive analysis
of the SUSY contributions to the hadronic EDMs will be discussed
elsewhere.
}

In general SUSY models, both left-handed and right-handed squarks have
flavor mixings. In this case, the CEDM of the $i$-th light quark $q_i$
is generated by a diagram in Fig.~2(a), and it can be enhanced by
$m_{q_j}/m_{q_i}$ when $j>i$, $(i,j=1,2,3)$. Using the mass insertion
technique, it is given by
\begin{eqnarray}
 {\widetilde d}_{q_i} = c\frac{\alpha_s}{4\pi}\frac{m_{\tilde g}}{m^2_{\tilde q}}
\lsp -\frac{1}{3}N_1(x) -3 N_2(x)\rsp {\mathrm Im}
\left[J_{ij}^{(q)}\right],
\label{SUSYEDM}
\end{eqnarray}
where $m_{\tilde{g}}$ and $m_{\tilde{q}}$ are the gluino and
averaged squark masses and $c$ is the QCD correction,
$c\sim 0.9$.  
The functions $N_i$ are given as
\begin{eqnarray}
N_1(x)&=&\frac{3+44x-36x^2-12x^3+x^4+12x(2+3x)\log x}{6(x-1)^6},
\\
N_2(x)&=&-  \frac{ 10 + 9x -18 x^2-x^3+3(1+6x+3x^2) \log x}{3(x-1)^6}.
\end{eqnarray}
The flavor violation in the squark mass terms contributes to the quark
EDMs via combinations of $J_{ij}^{(q)}$,
\begin{eqnarray}
J_{ij}^{(q)}&\equiv&( \delta_{LL}^{(q)})_{ij}
\, (\delta_{LR}^{(q)})_{jj}\,(\delta_{RR}^{(q)})_{ji}.
\end{eqnarray}
The mass insertion parameters,
$(\delta_{LL}^{(q)})_{ij}$,
$(\delta_{RR}^{(q)})_{ij}$ and $( \delta_{LR}^{(q)})_{ij}$,
 are defined as 
\begin{eqnarray}
(\delta_{LL}^{(q)})_{ij}=
\frac{\left(m_{\tilde{q}_L}^2\right)_{ij}}{m^2_{\tilde{q}}},&&
(\delta_{RR}^{(q)})_{ij}=
\frac{\left(m_{\tilde{q}_R}^2\right)_{ij}}{m^2_{\tilde{q}}},
\end{eqnarray}
for $q=u,d$, and
\begin{eqnarray}
( \delta_{LR}^{(d)})_{ij} 
= \delta_{ij} \frac{m_{d_j}\left(A_j^{(d)} -\mu\tan\beta\right)}{m^2_{\tilde{d}}},
&&
( \delta_{LR}^{(u)})_{ij} 
= \delta_{ij} \frac{m_{d_j}\left(A_j^{(u)} -\mu\cot\beta\right)}{m^2_{\tilde{u}}}.
\end{eqnarray}
Here, $(m_{\tilde{q}_{L(R)}}^2)$ is the left-handed (right-handed)
squark mass matrix. Here, we assume that the flavor-conserving 
SUSY breaking terms do not have the CP phases since they are 
already strongly constrained, and we keep terms induced by
the off-diagonal terms in the squark mass matrices

In the typical SUSY models, the left-handed squark mixings are
governed by the CKM matrix, and even if the universal scalar mass
hypothesis is imposed at high energy scale, the radiative correction
induces the off-diagonal terms as
\begin{eqnarray}
(\delta_{LL}^{(d)})_{12}=O(\lambda^5)\simeq 3\times 10^{-4},\;
&&
(\delta_{LL}^{(d)})_{23}=O(\lambda^2)\simeq 4\times 10^{-2},\;
\nonumber\\
(\delta_{LL}^{(d)})_{13}=O(\lambda^3)\simeq 8\times 10^{-3}, 
&&
\label{lambdarg}
\end{eqnarray}
for $\lambda\sim 0.2$.  On the other hand, the right-handed squark
mixing is rather model dependent.  In the SU(5) SUSY GUT with
right-handed neutrinos, large right-handed squark mixings can be
induced by the neutrino Yukawa couplings \cite{Moroi:2000tk}. They are
constrained from the $K^0$-$\overline{K^0}$, $D^0$-$\overline{D^0}$,
$B^0$-$\overline{B^0}$ mixings experiments \cite{Gabbiani:1996hi};
\begin{eqnarray}
\left|(\delta_{RR}^{(d)})_{12}\right| &\lsim& 4\times 10^{-2},
\nonumber\\
\left|(\delta_{RR}^{(u)})_{12}\right| &\lsim& 1 \times 10^{-1},
\nonumber\\
\left|(\delta_{RR}^{(d)})_{13}\right| &\lsim& 1 \times 10^{-1},
\label{kdb}
\end{eqnarray}
for $m_{\tilde q}=m_{\tilde g}=500$ GeV.  

The quark CEDMs are estimated  from Eq.~(\ref{SUSYEDM}) as
\begin{eqnarray}
  e {\tilde d}_u
&=&- 7.9 \times 10^{-28} \sin\theta_u^{(2)}\, {e\, {\rm cm}} \,\times
\nonumber\\
&&
\left(\frac{m_{\tilde{q}}}{500\mathrm{GeV}}\right)^{-3}
\left(\frac{(\delta_{LL}^{(u)})_{12}}{3\times 10^{-4}}\right)
\left(\frac{(\delta_{RR}^{(u)})_{21}}{0.1}\right)
\left(\frac{A_c-\mu \cot\beta}{500\mathrm{GeV}}\right)
\nonumber~~~~\\
&-& 5.7 \times 10^{-24} \sin\theta_u^{(3)}\, {e\, {\rm cm}} \,\times
\nonumber\\
&&
\left(\frac{m_{\tilde{q}}}{500\mathrm{GeV}}\right)^{-3}
\left(\frac{(\delta_{LL}^{(u)})_{13}}{8\times 10^{-3}}\right)
\left(\frac{(\delta_{RR}^{(u)})_{31}}{0.1}\right)
\left(\frac{A_t-\mu \cot\beta}{500\mathrm{GeV}}\right),
~~~~\\
  e {\tilde d}_d
&=& -4.0 \times 10^{-28} \sin\theta_d^{(2)}\, {e\, {\rm cm}} \,\times
\nonumber\\
&&
\left(\frac{m_{\tilde{q}}}{500\mathrm{GeV}}\right)^{-3}
\left(\frac{(\delta_{LL}^{(d)})_{12}}{3\times 10^{-4}}\right)
\left(\frac{(\delta_{RR}^{(d)})_{21}}{4\times 10^{-2}}\right)
\left(\frac{\mu \tan\beta}{5000\mathrm{GeV}}\right)
\nonumber~~~~\\
&-&9.6 \times 10^{-25} \sin\theta_d^{(3)}\, {e\, {\rm cm}} \,\times
\nonumber\\
&&
\left(\frac{m_{\tilde{q}}}{500\mathrm{GeV}}\right)^{-3}
\left(\frac{(\delta_{LL}^{(d)})_{13}}{8\times 10^{-3}}\right)
\left(\frac{(\delta_{RR}^{(d)})_{31}}{0.1}\right)
\left(\frac{\mu \tan\beta}{5000\mathrm{GeV}}\right),
~~~~\\
  e {\tilde d}_s
&=& -4.8 \times 10^{-24} \sin\theta_s^{(3)}\, {e\, {\rm cm}} \,\times
\nonumber\\
&&
\left(\frac{m_{\tilde{q}}}{500\mathrm{GeV}}\right)^{-3}
\left(\frac{(\delta_{LL}^{(d)})_{23}}{0.04}\right)
\left(\frac{(\delta_{RR}^{(d)})_{32}}{0.1}\right)
\left(\frac{\mu \tan\beta}{5000\mathrm{GeV}}\right),~~~~
\end{eqnarray}
where we take $m_{\tilde{q}}=m_{\tilde{g}}$ and $\theta_{q_i}^{(j)}$ is the phase
of the SUSY parameters, {\it i.e.}
$\theta_{q_i}^{(j)}={\mathrm{arg}}[J_{ij}^{(q)}]$.  Here, we take the
experimental bounds for mass insertion parameters of the right-handed
squarks in Eq.~(\ref{kdb}) while the left-handed ones are given by
Eq.~(\ref{lambdarg}). We have neglected the $A$-terms in $\tilde{d}_d$
and $\tilde{d}_s$ since they are subdominant.

When one of the quark CEDMs saturates the hadronic EDM bounds,
we can obtain the following bounds on the SUSY CP phases
from the $^{199}$Hg atomic (neutron)  EDM experiments,\footnote{
Here we do not consider the quark EDM contributions to the neutron
EDM.  We find that they are subdominant compared with those of the
quark CEDMs in our calculation.
}
\begin{eqnarray}
\left| \sin\theta_u^{(2)}\right|&<& 30\, (47)~
\left(\frac{(\delta_{LL}^{(u)})_{12}}{3\times 10^{-4}}\right)^{-1}
\left(\frac{(\delta_{RR}^{(u)})_{21}}{0.1}\right)^{-1}
\left(\frac{A_c-\mu \cot\beta}{500\mathrm{GeV}}\right)^{-1},
~~~~\\
\left| \sin\theta_u^{(3)}\right|&<& 4.2\, (6.5)\times 10^{-3}~
\left(\frac{(\delta_{LL}^{(u)})_{13}}{8\times 10^{-3}}\right)^{-1}
\left(\frac{(\delta_{RR}^{(u)})_{31}}{0.1}\right)^{-1}
\left(\frac{A_t-\mu \cot\beta}{500\mathrm{GeV}}\right)^{-1},
~~~~\\
\left| \sin\theta_d^{(2)}\right|&<& 60\, (113)~
\left(\frac{(\delta_{LL}^{(d)})_{12}}{3\times 10^{-4}}\right)^{-1}
\left(\frac{(\delta_{RR}^{(d)})_{21}}{4\times 10^{-2}}\right)^{-1}
\left(\frac{\mu \tan\beta}{5000\mathrm{GeV}}\right)^{-1},
~~~~\\
\left| \sin\theta_d^{(3)}\right|&<& 2.5\, (4.7)\times 10^{-2}~
\left(\frac{(\delta_{LL}^{(d)})_{13}}{8\times 10^{-3}}\right)^{-1}
\left(\frac{(\delta_{RR}^{(d)})_{31}}{0.1}\right)^{-1}
\left(\frac{\mu \tan\beta}{5000\mathrm{GeV}}\right)^{-1},
~~~~\\
\left| \sin\theta_s^{(3)}\right|&<& 0.98\, (0.048)
\left(\frac{(\delta_{LL}^{(d)})_{23}}{0.04}\right)^{-1}
\left(\frac{(\delta_{RR}^{(d)})_{32}}{0.1}\right)^{-1}
\left(\frac{\mu \tan\beta}{5000\mathrm{GeV}}\right)^{-1},
\label{kkk}
\end{eqnarray}
where we take $m_{\tilde{q}}=m_{\tilde{g}}=500$GeV. The above bounds
on $\theta_u^{(2)}$ and $\theta_d^{(2)}$ are looser than than the
stringent $K^0$-$\overline{K}^0$ constraint. On the other hand, the CP
phases related to the 1-3 and 2-3 mixing angles are constrained by the
hadronic EDMs significantly. These bounds are expected to be improved
furthermore by the deuteron EDM measurements. 
\subsection{Hadronic EDMs and CP asymmetry in $B\to\phi K_S$}
\label{sec:phiks}
Let us consider a correlation between ${\tilde d}_s$ and $S_{\phi K_s}$
in the SUSY models. 
In Ref.~\cite{HS}, we have shown that there is a strong correlation
between them when both left-handed and right-right handed squarks have
flavor mixings. In such a case, the dominant
contribution to $S_{\phi K_s}$
is supplied by a diagram with the double
mass insertion of $(\delta_{RR}^{(d)})_{32}$ and
$(\delta_{RL}^{(d)})_{33}$ (Fig.~2(b)). 
The contribution of Fig.2~(b) is given as
\begin{eqnarray}
 H&=& - C_8^{R} \frac{g_s}{8\pi^2}
    m_b\overline{s_R}(G\sigma) b_L,
\end{eqnarray}
where
\begin{eqnarray}
 C_8^{R}&=&\frac{\pi \alpha_s}{m^2_{\tilde{q}}}
\frac{m_{\tilde{g}}}{m_b}
(\delta_{LR}^{(d)})_{33}(\delta_{RR}^{(d)})_{32}
(-\frac{1}{3} M_1(x)-3 M_2(x)),
\end{eqnarray}
up to the QCD correction. Here,
\begin{eqnarray}
 M_1(x)&=&\frac{1+9x-9x^2-x^3+ (6x+6x^2)\log x}{2(x-1)^5},
\\
 M_2(x)&=&-\frac{3-3x^2+(1+4x+x^2)\log x}{(x-1)^5}.
\end{eqnarray}
In a limit of $x\rightarrow 1$, $C_8^{R}$ is reduced to
\begin{eqnarray}
 C_8^{R}&=&\frac{7\pi \alpha_s}{60{m_b} m_{\tilde{q}}}
(\delta_{LR}^{(d)})_{33}
(\delta_{RR}^{(d)})_{32}.
\label{c8ap}
\end{eqnarray}
Using Eqs. (\ref{SUSYEDM}) and (\ref{c8ap}),
we find a strong correlation between ${\widetilde d}_s$ and $C_8^R$ as
\begin{eqnarray}
{\widetilde d}_s &=& -\frac{m_b}{4\pi^2} \frac{11}{21}
{\mathrm{Im}}\left[( \delta_{LL}^{(d)})_{23} C_8^{R}\right]
\label{massin}
\end{eqnarray}
up to the QCD correction. The coefficient $11/21$ in Eq.~(\ref{massin})
changes from 1 to $1/3$ for $0<x<\infty$.

In Fig.~3, we show the correlation between $\widetilde{d}_s$ and $S_{\phi K_s}$
assuming a relation $\widetilde{d}_s = -{m_b}/(4\pi^2) (11/21){\mathrm{Im}}[(
\delta_{LL}^{(d)})_{23}C_8^{R}]$ up to the QCD correction. Here, we
take $(
\delta_{LL}^{(d)})_{23} =-0.04$, ${\mathrm{arg}}[C_8^{R}]=\pi/2$ and 
$|C_8^R|$ corresponding to $10^{-5}<|(\delta_{RR}^{(d)})_{32}|<0.5$. The
matrix element of chromomagnetic moment in $B\rightarrow \phi K_s$ is
\begin{eqnarray}
\langle \phi K_S|\frac{g_s}{8\pi^2}m_b(\bar{s}_i \sigma^{\mu\nu}T^a_{ij}P_Rb_j)
G^a_{\mu\nu}| \overline{B}_d\rangle &=&
\kappa \frac{4\alpha_s}{9\pi}(\epsilon_\phi p_B)f_\phi m_\phi^2 F_+(m_\phi^2),
\label{kappa}
\end{eqnarray}
and $\kappa=-1.1$ in the heavy-quark effective theory \cite{Harnik:2002vs}.
Since $\kappa$ may suffer from the large hadron uncertainty, we show
the results for $\kappa=-1$ and $-2$. 
From this figure, the deviation
of $S_{\phi K_s}$ from the SM prediction due to the gluon penguin
contribution should be suppressed when  the constraints on $\widetilde{d}_s$ 
from the $^{199}$Hg atomic and the neutron EDMs are applied. 
Comparing Fig.~2 in Ref. \cite{HS}, the $^{199}$Hg atomic EDM bound
allows a sizable deviation of $S_{\phi K_s}$, especially in cases of 
$\kappa=-2$ and $|(\delta_{LL}^{(d)})_{23}|$ smaller than 0.04.
It comes from the new theoretical estimation of the strange CEDM constraint
in Eq.~(\ref{ds}) and the numerical error in previous calculation.
However, we find that the neutron EDM gives a strong
bound on $S_{\phi K_s}$. Moreover, $S_{\phi K_s}$ may be constrained
further by the future deuteron EDM measurements.
Therefore, the hadronic EDMs give a very important implication
to $S_{\phi K_s}$.

\section{Conclusion and discussion}
\label{sec:summary}

We have considered hadronic CP violation induced by chromoelectric
dipole moments of light quarks and the QCD theta parameter, especially
paying an attention to the strange quark CEDM.  First, we have derived
the effective CP violating nucleon interactions induced by the CEDMs
and the theta parameter using the chiral Lagrangian technique. In order to take
into account the strange quark contributions, we have used the SU(3)
chiral Lagrangian. Using the effective CP violating Lagrangian, we
calculated the EDMs of the $^{199}$Hg atom, neutron and deuteron.

The $^{199}$Hg atomic EDM is sensitive to the CP violating nuclear force
induced by the $\pi$ and $\eta^0$ exchange diagrams. Though the
contribution of strange quark EDM is evaluated from the eta exchange
diagrams in the previous papers, it is found in the new evaluation of
the Shiff momentum that the isoscalar channel contribution in the
$\pi$ and $\eta^0$ exchange is suppressed. We found that the isospin
breaking nucleon, interactions originated from the $\pi^0$-$\eta^0$
mixing, leads to the similar constraint on the strange quark CEDM to
the previous one.

We evaluated the strange quark CEDM contribution to the neutron EDM
using the standard meson loop calculation.  This is originated from
the one loop diagram involving the $K$ meson. We have found that the
neutron EDM gives a stronger bound on the strange quark CEDM in this
calculation than the current $^{199}$Hg atomic EDM experiment, since the
contribution to the neutron EDM is suppressed by the loop factor at
most. While this calculation has theoretical uncertainties, it also
suggests that the strange quark CEDM should be small.

The new technique for the measurement of the deuteron EDM has a great
impact on the strange quark CEDM if it is realized. If they establish
the sensitivity of $d_D \sim  10^{-27}e\,cm$, we may probe the
new physics to the level of $e\tilde{d}_s\sim 10^{-26}\;e\;cm$, which 
is stronger than the bound from the neutron EDM.

In order to demonstrate an implication of our result on the SUSY
models, we calculate the gluino contributions to the quark CEDMs with
the flavor violating mass insertions. It is usually considered that
the EDMs are sensitive to the flavor diagonal CP phases.  However,
when both left-handed and right-handed quark mixing exist, the CEDMs
can be enhanced by the left-right squark mixings. Since the typical
SUSY models have the left-handed squark mixing, the EDMs can give
strong constraints on the flavor dependent SUSY phases.  These
constraints on the SUSY phases can give important implications to
other SUSY phenomenology, including the $B$ physics.
As an example, we have show that there is a strong correlation 
between the strange quark CEDM and $S_{\phi K_S}$.
The current bounds on the strange quark CEDM from $^{199}$Hg atomic and the neutron
EDMs imply that the deviation of $S_{\phi K_S}$ from the SM should
be strongly suppressed. 

\section*{Acknowledgments}
We would like to appreciate useful discussion with Dr M.~Kakizaki and
Dr M.~Nagai.  The work of J.H. is supported in part by the
Grant-in-Aid for Science Research, Ministry of Education, Science and
Culture, Japan (No.15540255, No.13135207 and No.14046225).  Also, the
work of Y.S. is supported in part by the 21st century COE program,
``Exploring New Science by Bridging Particle-Matter Hierarchy''.

\appendix
\section{CP even terms in chiral Lagrangian}

In the chiral Lagrangian the interaction can be expanded by
derivatives and the lowest order terms are important at the low
energy. Here, we summarize the relevant CP even terms in the chiral
Lagrangian.  The relevant CP conserving couplings between the nucleons
and the pseudoscalars are
\begin{eqnarray}
 i{\cal L}_p & \equiv& g_{\pi^c N_aN_b}\, \overline{N_a}\gamma_5 N_b \pi^c
\nonumber\\
&=&
\frac{m_p}{f_\pi}(D+F) \overline{p}\gamma_5 p \pi^0
- 
\frac{m_n}{f_\pi}(D+F) \overline{n}\gamma_5 n \pi^0
+
\frac{m_p+m_n}{\sqrt{2}f_\pi}(D+F) \overline{n}\gamma_5 p \pi^-
\nonumber\\
&-&
\frac{m_p}{\sqrt{3}f_\pi}(D-3F) \overline{p}\gamma_5 p \eta^0
-
\frac{m_n}{\sqrt{3}f_\pi}(D-3F)\overline{n}\gamma_5 n \eta^0
\nonumber\\
&+&
\frac{m_\Sigma+m_p}{\sqrt{2}f_\pi}(D-F)
\overline{\Sigma^+}\gamma_5 p \overline{K^0}
-
\frac{m_\Sigma+m_n}{{2}f_\pi}(D-F) 
\overline{\Sigma^0}\gamma_5 n \overline{K^0}
\nonumber\\
&-&
\frac{m_\Lambda+m_n}{2\sqrt{3}f_\pi}(D+3F) 
\overline{\Lambda^0}\gamma_5 n \overline{K^0}
+
\frac{m_\Sigma+m_p}{{2}f_\pi}(D-F) \overline{\Sigma^0}\gamma_5 p K^-
\nonumber\\
&-&\frac{m_\Lambda+m_p}{2\sqrt{3}f_\pi}(D+3F)
\overline{\Lambda^0}\gamma_5 p {K^-}
+
\frac{m_\Sigma+m_n}{\sqrt{2}f_\pi}(D-F)\overline{\Sigma^-}\gamma_5 n {K^-},
\end{eqnarray}
where we have used the equation of motion for the nucleon fields,
$\overline{N}\gamma_\mu\gamma_5 N'\partial^\mu M=-i(m_N+m_{N'})
\overline{N}\gamma_5 N' M$. The coupling constants are given by
numerically,
\begin{eqnarray}
 &&g_{\pi pp}=12.6,~ g_{\pi nn}=-12.6,~ g_{\pi pn}=17.9,~\\
 &&g_{\eta pp}=2.98,~ g_{\eta nn}=2.98,~ g_{K \Sigma n}=5.98,~
\end{eqnarray}
for $D=0.81$ and  $F=0.44$. 

The pseudoscalars are massless due to the SU(3)$_L\times$SU(3)$_R$ symmetry.
However, the SU(3)$_L\times$SU(3)$_R$ symmetry is violated by small 
quark masses and the pseudoscalars become massive. 
When the quark masses $m=diag(m_u,m_d,m_s)$ are taken into account,
we can introduce the following terms,
\begin{eqnarray}
\label{L1}
 {\cal L}_1 &=& v^3 \tr \left(U^\dagger m + m U \right)
\nonumber\\
       &+& a_1 \tr\left(\overline{B}\left(\xi^\dagger m \xi^\dagger 
				   + \xi m\xi\right)B\right)
       + a_2 \tr\left(\overline{B}B\left(\xi^\dagger m \xi^\dagger 
				   + \xi m\xi\right)\right)
\nonumber\\
       &+& b_1 \tr\left(\overline{B}\gamma_5\left(\xi^\dagger m \xi^\dagger 
				   - \xi m\xi\right)B\right)
       + b_2 \tr\left(\overline{B}\gamma_5B\left(\xi^\dagger m \xi^\dagger 
				   - \xi m\xi\right)\right)
\nonumber
\\
 &=& -2 \frac{v^3}{f_\pi^2}\tr\left(M^2 m\right)
+2 a_1 \tr\left(\overline{B}m B\right)
+2 a_2 \tr\left(\overline{B} B m\right)
\nonumber\\
&-& \frac{2ib_1}{\sqrt{2}f_\pi}
\tr\left(\overline{B}\gamma_5(M m+ mM)B\right)
-\frac{2ib_2}{\sqrt{2}f_\pi}
\tr\left(\overline{B}\gamma_5 B(M m+ mM)\right)+...\;\;.
\end{eqnarray}
In the estimation of the hadronic EDMs, the
isospin breaking effect is important as shown in text.
The isospin symmetry is violated by the quark mass term and 
the $\pi^0$-$\eta^0$ mixing occur through the mass terms.
From Eq.~(\ref{L1}), the $\pi^0$-$\eta^0$ mass matrix is given by
\begin{eqnarray}
\label{pieta}
{\cal L_{\pi-\eta}}= \frac{2v^3}{f_\pi^2}
\left(\pi, \eta \right) 
 \left(
\begin{array}{cc}
m_u+m_d& \frac{1}{\sqrt{3}}(m_u-m_d)\\
\frac{1}{\sqrt{3}}(m_u-m_d)&\frac{1}{3}({m_u+m_d+4m_s})
\end{array}
\right)
 \left(
\begin{array}{c}
 \pi\\
 \eta
\end{array}
\right).
\end{eqnarray}
The mass eigenstates are defined by
\begin{eqnarray}
 \left(
\begin{array}{c}
 \pi'\\
 \eta'
\end{array}
\right) =
 \left(
\begin{array}{cc}
 \cos\theta& \sin\theta\\
 -\sin\theta& \cos\theta
\end{array}
\right)
 \left(
\begin{array}{c}
 \pi\\
 \eta
\end{array}
\right),
\end{eqnarray}
where $\tan\theta=-\sqrt{3}(m_u-m_d)/(4m_s)\simeq 0.01$
for $m_u=5.1$ MeV, $m_d=9.3$ MeV, $m_s=175$ MeV.

\section{CP-odd terms in chiral Lagrangian}

From the effective Lagrangian in Eq.~(\ref{NNCP2}), the CP violating
nucleon interaction terms are written as follows. Here, we do not
include the $\pi^0$-$\eta^0$ mixing.
\begin{eqnarray}
 {\cal L}_{\rm {\sla CP}}
&\equiv& \overline{g}_{\pi_cN^aN_b} \overline{N}_aN_b\pi^c
\nonumber\\
&=&
\frac{1}{f_\pi}\left(A_u \langle\overline{u}u\rangle 
-A_d\langle \overline{d}d \rangle \right)\overline{p}p\pi^0
\nonumber\\
&+&
\frac{1}{f_\pi}\left(A_u \langle\overline{d}d\rangle 
-A_d\langle \overline{u}u \rangle \right)\overline{n}n\pi^0
\nonumber\\
&+&
\frac{1}{\sqrt{2}f_\pi}\left(A_u +A_d\right)\left( \langle\overline{u}u\rangle 
-\langle \overline{d}d \rangle \right)\overline{n}p\pi^-
\nonumber\\
&+&\frac{1}{\sqrt{3}f_\pi}
\left(A_u \langle\overline{u}u\rangle 
+A_d\langle \overline{d}d \rangle
-2A_s\langle \overline{s}s \rangle
 \right)\overline{p}p\eta^0
\nonumber\\
&+&\frac{1}{\sqrt{3}f_\pi}
\left(A_u \langle\overline{d}d\rangle 
+A_d\langle \overline{u}u \rangle
-2A_s\langle \overline{s}s \rangle
 \right)\overline{n}n\eta^0
\nonumber\\
&-&
\frac{1}{\sqrt{2}f_\pi}\left(A_d +A_s\right)\left(\langle\overline{d}d\rangle 
-\langle \overline{s}s \rangle \right)\overline{\Sigma^+}p\overline{K^0}
\nonumber\\
&+&
\frac{1}{{2}f_\pi}\left(A_d +A_s\right)\left(\langle\overline{d}d\rangle 
-\langle \overline{s}s \rangle \right)\overline{\Sigma^0}n \overline{K^0}
\nonumber\\
&+&
\frac{1}{2\sqrt{3}f_\pi}\left(A_d +A_s\right)
\left(\langle\overline{d}d\rangle 
+\langle \overline{s}s \rangle 
-2\langle \overline{u}u \rangle 
\right)\overline{\Lambda^0}n \overline{K^0}.
\nonumber\\
&-&
\frac{1}{{2}f_\pi}\left(A_u +A_s\right)\left(\langle\overline{d}d\rangle 
-\langle \overline{s}s \rangle \right)\overline{\Sigma^0}p {K^-}
\nonumber\\
&+&
\frac{1}{2\sqrt{3}f_\pi}\left(A_u +A_s\right)
\left(\langle\overline{d}d\rangle 
+\langle \overline{s}s \rangle 
-2\langle \overline{u}u \rangle 
\right)\overline{\Lambda^0}p {K^-}
\nonumber \\
&-&
\frac{1}{\sqrt{2}f_\pi}\left(A_u +A_s\right)\left(\langle\overline{d}d\rangle 
-\langle \overline{s}s \rangle \right)\overline{\Sigma^-}n{K^-}.
\end{eqnarray}

\newpage
%
%
\newcommand{\Journal}[4]{{\sl #1} {\bf #2} {(#3)} {#4}}
\newcommand{\APJ}{Ap. J.}
\newcommand{\CJP}{Can. J. Phys.}
\newcommand{\MPL}{Mod. Phys. Lett.}
\newcommand{\NC}{Nuovo Cimento}
\newcommand{\NP}{Nucl. Phys.}
\newcommand{\PL}{Phys. Lett.}
\newcommand{\PR}{Phys. Rev.}
\newcommand{\PRep}{Phys. Rep.}
\newcommand{\PRL}{Phys. Rev. Lett.}
\newcommand{\PTP}{Prog. Theor. Phys.}
\newcommand{\SJNP}{Sov. J. Nucl. Phys.}
\newcommand{\ZP}{Z. Phys.}

\begin{figure}[p]
\label{fig:dn}
\begin{picture} (550,160)(0,0)
\SetOffset(-40,0)
\DashArrowArcn(135,25)(75,180,0){3}
\Text(135,120)[]{$\pi^-$}

\ArrowLine(30,25)(60,25)
\ArrowLine(60,25)(210,25)
\ArrowLine(210,25)(240,25)
\Vertex(60,25){3}

\Text(45,15)[]{$n$}
\Text(135,15)[]{$p$}
\Text(225,15)[]{$n$}
\Photon(172,110)(195,135){2}{5}
\Text(203,145)[]{$\gamma$}
\SetOffset(200,0)
\DashArrowArc(135,25)(75,0,180){3}
\Text(135,120)[]{$K^-$}

\ArrowLine(30,25)(60,25)
\ArrowLine(60,25)(210,25)
\ArrowLine(210,25)(240,25)
\Vertex(60,25){3}

\Text(45,15)[]{$n$}
\Text(135,15)[]{$\Sigma$}
\Text(225,15)[]{$n$}
\Photon(172,110)(195,135){2}{5}
\Text(203,145)[]{$\gamma$}
\end{picture}

\caption{ One loop diagrams for the neutron EDM.
The blob represents the CP violating coupling.
}

\end{figure}
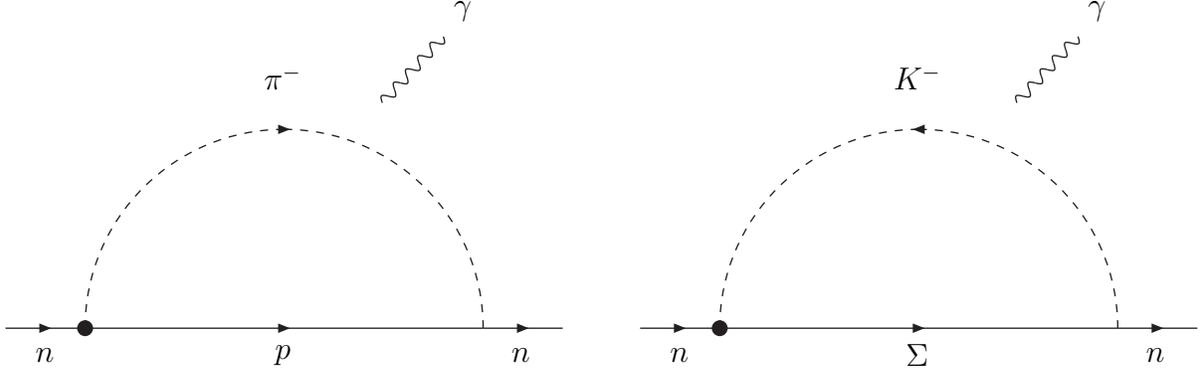

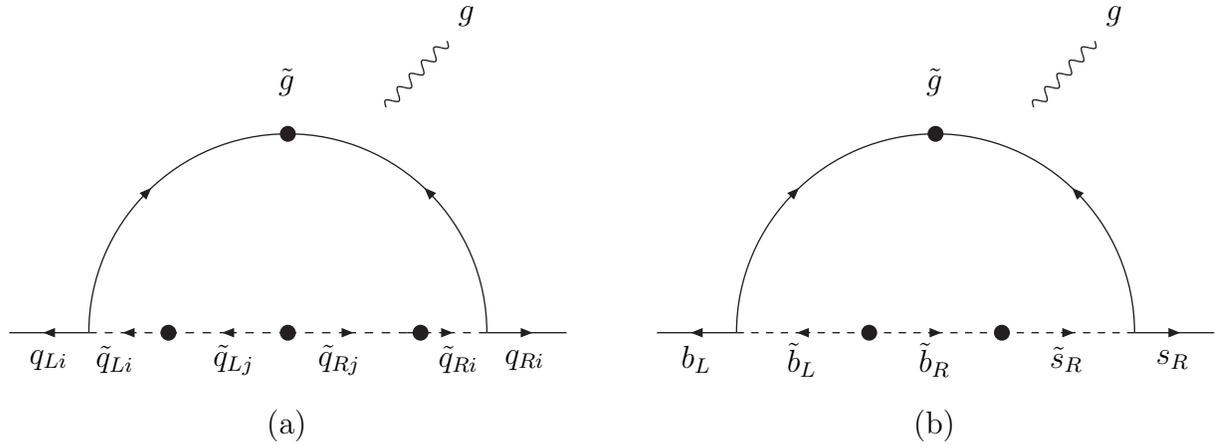
\begin{figure}[p]
\label{fig:CEDM}
\begin{center} 
\begin{picture}(455,140)(30,-20)

\ArrowArcn(135,25)(75,180,90)
\ArrowArc(135,25)(75,0,90)
\Vertex(135,100){3}
\Text(135,120)[]{$\tilde{g}$}

\ArrowLine(60,25)(30,25)
\DashArrowLine(90,25)(60,25){3}             \Vertex(90,25){3}
\DashArrowLine(135,25)(90,25){3}             \Vertex(135,25){3}
\DashArrowLine(135,25)(180,25){3}       \Vertex(185,25){3}  
\DashArrowLine(180,25)(210,25){3}        
\ArrowLine(210,25)(240,25)

\Text(45,15)[]{$q_{Li}$}
\Text(70,15)[]{$\tilde{q}_{Li}$}
\Text(115,15)[]{$\tilde{q}_{Lj}$}
\Text(155,15)[]{$\tilde{q}_{Rj}$}
\Text(200,15)[]{$\tilde{q}_{Ri}$}
\Text(225,15)[]{$q_{Ri}$}

\Photon(172,110)(195,135){2}{5}
\Text(203,145)[]{$g$}
\Text(135,-10)[]{(a)}

\SetOffset(245,0)

\ArrowArcn(135,25)(75,180,90)
\ArrowArc(135,25)(75,0,90)
\Vertex(135,100){3}
\Text(135,120)[]{$\tilde{g}$}

\ArrowLine(60,25)(30,25)
\DashArrowLine(110,25)(60,25){3}             \Vertex(110,25){3}
\DashArrowLine(110,25)(160,25){3}            \Vertex(160,25){3}        
\DashArrowLine(160,25)(210,25){3}           
\ArrowLine(210,25)(240,25)

\Text(45,15)[]{$b_L$}
\Text(85,15)[]{$\tilde{b}_L$}
\Text(135,15)[]{$\tilde{b}_R$}
\Text(185,15)[]{$\tilde{s}_R$}
\Text(225,15)[]{$s_R$}

\Photon(172,110)(195,135){2}{5}
\Text(203,145)[]{$g$}

\Text(135,-10)[]{(b)}

\end{picture} 

\caption{(a) The dominant diagram contributing to the CEDMs of light
quarks when both the left-handed and right-handed squarks have 
flavor mixings.
(b) The dominant SUSY diagram contributing to the CP asymmetry in
$B\rightarrow \phi K_s$ when the right-handed squarks have a mixing. }

\end{center}
\end{figure}

\begin{figure}
 \centerline{
\epsfxsize = 0.8\textwidth \epsffile{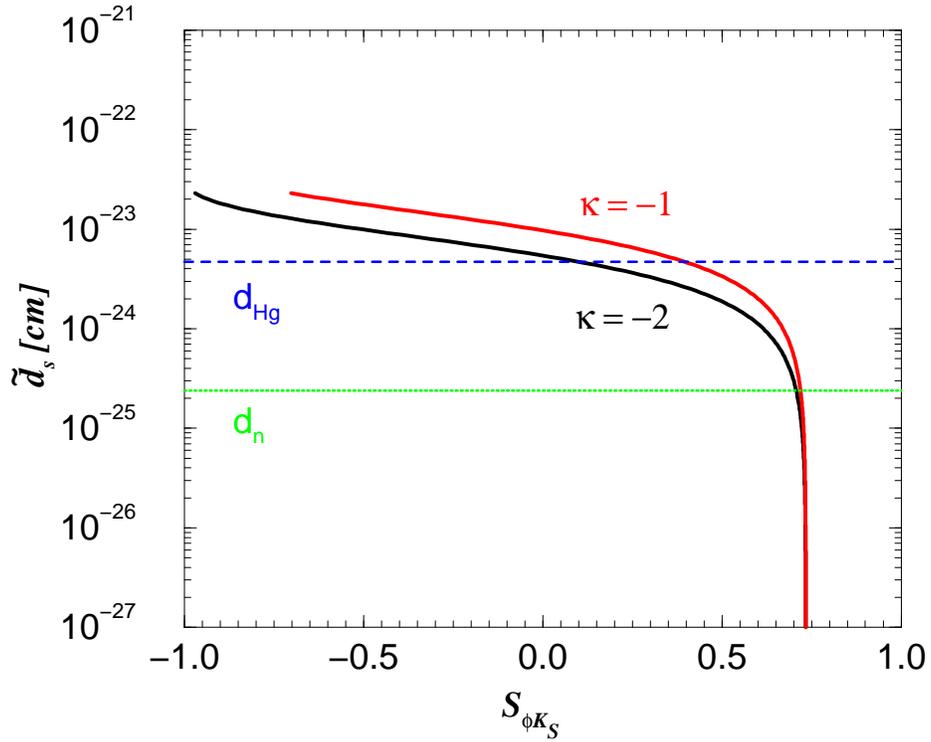} 
}
\vspace*{-5mm}
\caption{
The correlation between ${\widetilde d}_s$ and $S_{\phi K_s}$ assuming
${\widetilde d}_s = -{m_b}/(4\pi^2)(11/21)
{\mathrm{Im}}[ ( \delta_{LL}^{(d)})_{23}C_8^{R}]$. Here, $( \delta_{LL}^{(d)})_{23}
=-0.04$ and ${\mathrm{arg}}[C_8^{R}]=\pi/2$. $\kappa$ comes from 
the matrix element of chromomagnetic
moment  in $B\rightarrow \phi K_s$.
The dashed (dotted) line is the upperbound on ${\widetilde d}_s$ from the EDM of 
$^{199}$Hg atom (neutron). 
}
\end{figure}
\end{document}